\documentclass[pss]{wiley2sp} 
\usepackage{amssymb}
\usepackage{amsmath}
\usepackage{bm}     
\newcommand{\ua}{\uparrow}
\newcommand{\da}{\downarrow}
\newcommand{\eq}{\begin{equation}}
\newcommand{\eqx}{\end{equation}}
\newcommand{\eqn}{\begin{eqnarray}}
\newcommand{\eqnx}{\end{eqnarray}}

\newcommand{\veck}{{\bf k}}
\newcommand{\veci}{{\bf i}}
\newcommand{\vecj}{{\bf j}}

\newcommand{\vecl}{{\bf l}}
\newcommand{\vecm}{{\bf m}}
\newcommand{\veco}{{\bf 0}}
\begin{document}

\title{Evaluation techniques for Gutzwiller wave functions in finite dimensions}

\titlerunning{Evaluation techniques for Gutzwiller wave functions}

\author{Jan Kaczmarczyk\textsuperscript{\textsf{\bfseries 1},\textsf{\bfseries 2}}, 
Tobias Schickling\textsuperscript{\textsf{\bfseries 3}}, and 
J\"org B\"unemann\textsuperscript{\Ast,\textsf{\bfseries 3},\textsf{\bfseries 4}}}

\authorrunning{J.\ Kaczmarczyk et al.}
\mail{e-mail
  \textsf{buenemann@gmail.com}}

\institute{%
  \textsuperscript{1}\,Institute of Science and Technology Austria, Am Campus 1, A-3400, Klosterneuburg, Austria\\
 \textsuperscript{2}\,Marian Smoluchowski Institute of Physics, Jagiellonian
 University, \L{}ojasiewicza 11, 30-348 Krak\'ow, Poland\\
  \textsuperscript{3}\,Fachbereich Physik, Philipps Universit\"at, 
Renthof 6, 35032 Marburg, Germany\\
\textsuperscript{4}\,Institut f\"ur Physik, BTU Cottbus-Senftenberg, 
P.O.\ Box 101344, 03013 Cottbus, Germany}

\received{XXXX, revised XXXX, accepted XXXX} 
\published{XXXX} 


\abstract{
\abstcol{We give a comprehensive introduction into a diagrammatic method that allows for the 
 evaluation of Gutzwiller wave functions in finite spatial dimensions. 
 We discuss in detail some numerical schemes that turned out to be useful in 
 the real-space evaluation of the diagrams. 
The method is applied to the problem of d-wave
 }{superconductivity in a two-dimensional 
 single-band Hubbard model. Here, we discuss in particular the role of 
long-range contributions in our diagrammatic expansion. We further reconsider our previous 
 analysis on the kinetic energy gain in the superconducting state.}
}

\maketitle

\section{Introduction}
For the study of the ground-state properties of quantum systems, 
variational wave functions can be a powerful tool.  The most
 common variational approach in the theory of correlated electron systems
 is the Hartree--Fock approximation (HFA) which is  
 based on (variational) single-particle product wave functions 
 $|\Psi_0 \rangle$. 
If applied to 
  systems with attractive, e.g., phonon-mediated interactions, the
 HF theory leads to the celebrated BCS theory on 
 superconductivity~\cite{bardeen1957}.

The crucial first step in any variational approach is the calculation 
 of the energy expectation value for a given class of variational 
 wave functions. For HF wave functions this can always be achieved
 by means of Wick's theorem, 
 which explains the popularity of this approach. 
Many phenomena in correlated electron systems, however, 
 cannot be described properly by the HFA, especially, 
 in (effectively) one- or two-dimensional systems. This holds, 
 in particular, for unconventional superconductivity
 which has been observed in a number of materials, such as 
 Cuprates, Ruthenates and iron-based Pnictides. 

A way to improve the  HFA is 
 based on `Jastrow wave functions' which have the form 
\cite{feenberg1969,clements1993}
\begin{equation}
|\Psi_{\rm J} \rangle=\hat{P}_{\rm J} |\Psi_0 \rangle\;.
\end{equation}
Here, $\hat{P}_{\rm J} $ is an operator which has been chosen 
 in various ways in the literature~\cite{gutzwiller1963,gutzwiller1964,gutzwiller1965,Yokoyama1,Yokoyama2,Baeriswyl,Hetenyi} and is meant to account for correlation 
 effects which are not captured by the Hartree--Fock (single-particle) wave function 
$|\Psi_0 \rangle$. One of the simplest examples for such a 
Jastrow wave function is the Gutzwiller wave 
function~\cite{gutzwiller1963,gutzwiller1964,gutzwiller1965} 
 which will be used in this work. 

Evaluating expectation values for 
Jastrow (or Gutzwiller) wave functions is a difficult many-particle 
 problem which, in general, can only by tackled by numerical 
 techniques, such as the `variational Monte-Carlo method' 
(VMC)~\cite{horsch1983,koch1999,edegger2007}. We have 
recently developed a diagrammatic scheme for the evaluation of  
  expectation values for  Gutzwiller wave functions. Unlike the 
 VMC, our method adresses the infinite systems and, hence, it does not 
 suffer from the typical finite-size errors of VMC.  
As shown in Ref.~\cite{buenemann2012a} , 
our approach allows us to study e.g., 
the stability of
 nematic (`Pomeranchuk') phases in two-dimensional Hubbard models. 
First results on the stability of 
 superconducting ground states in these models 
 have been presented in Ref.~\cite{buenemann2012b}.

 In this work we will give a comprehensive introduction into 
 the technical details of our approach for the 
 study of superconducting ground states and present numerical results
 which complement those published in previous work,  
Refs.~\cite{buenemann2012a,buenemann2012b,buenemann2014b,buenemann2014a,kaczmarczyk2014a}. We will introduce, in particular, a new  way to evaluate 
 diagrams which contain long-range correlations.

 Our
 presentation is organised as follows. In Section~\ref{se1} we introduce 
  the diagrammatic method which we use for the investigation
 of the single-band model. The class of diagrams which requires a 
 special treatment due to their long-range contributions  is discussed in 
 Section~\ref{app4}. In Section~\ref{se3} we show numerical results 
 and focus, in particular, on the convergence of our diagrammatic scheme.
 Our presentation is closed by a Summary and Outlook in  Section~\ref{se4}. 
 Some technical parts of the presentation are referred to three appendices

\section{Model and Method}\label{se1}
We investigate the single-band Hubbard model 
\begin{equation}
\hat{H}=\hat{H}_0 + U\sum_{\veci}  \hat{d}_{\veci}    \,, 
\label{eq:1}
\end{equation}
in two dimensions, where
\begin{equation}
\hat{H}_0=\sum_{\veci,\vecj,\sigma}t_{\veci,\vecj}\hat{c}_{\veci,\sigma}^{\dagger}
\hat{c}_{\vecj,\sigma}^{\phantom{\dagger}} 
\,\,\, , \,\,\,
 \hat{d}_{\veci}\equiv\hat{n}_{\veci,\uparrow}\hat{n}_{\veci,\downarrow}\, .
\label{eq:1uu}
\end{equation}
 Here, $\veci=(i_1,i_2)$ denotes one of 
the $L$~sites on a square lattice, and $\sigma=\uparrow,\downarrow$.  
The properties of this model will be studied in the thermodynamic limit 
 $L\to \infty$ by means of the 
 variational wave functions
\begin{equation}
|\Psi_{\rm G}\rangle=\hat{P}_{\rm G}|\Psi_0\rangle
=\prod_{\veci}\hat{P}_{\veci;{\rm G}}|\Psi_0\rangle \; ,
\label{eq:1.2}                                                  
\end{equation}
first introduced by 
Gutzwiller~\cite{gutzwiller1963}, 
where $|\Psi_0\rangle$ is a (normalised) single-particle 
product state and the
local `Gutzwiller correlator' is defined by
\begin{equation}\label{eq:1.2b} 
\hat{P}_{\veci}=\sum_{\Gamma}\lambda_{\Gamma}
|\Gamma \rangle_{\veci\,\veci}\! \langle \Gamma |\;.
\end{equation}
It contains the variational parameters 
$\lambda_{\Gamma}$ for the four local states 
\begin{equation}
|\Gamma\rangle_{\veci}
\in \left\{|\emptyset\rangle_{\veci}, |\uparrow\rangle_{\veci}, 
|\downarrow\rangle_{\veci}, |\uparrow\downarrow\rangle_{\veci}\right\}
\end{equation}
for the empty, singly, or doubly occupied site~$\veci$. Note 
that in Eq.~(\ref{eq:1.2b}) we have already assumed 
 a translationally invariant ground state which allows us to 
 work with parameters $\lambda_{\Gamma}$ that do not depend on 
 the lattice site $\veci$.     

 The single particle state~$|\Psi_0\rangle$ is also a variational 
 object and may be chosen as the ground state of an 
 effective single-particle Hamiltonian, 
\begin{equation}
\hat{H}_0^{\rm eff} =
\sum_{\veci,\vecj,\sigma}t^{\rm eff}_{\veci,\vecj}
\hat{c}_{\veci,\sigma}^{\dagger}\hat{c}_{\vecj,\sigma}^{\phantom{\dagger}}
+ \sum_{\veci \neq \vecj} \bigl( \Delta^{\rm eff}_{\veci,\vecj}
\hat{c}_{\veci,\ua}^{\dagger}\hat{c}_{\vecj,\da}^{\dagger} +
{\rm h.c.} \bigr) \label{eq:iou1}\;.
\end{equation}
The effective hopping and pairing parameters $t^{\rm eff}_{\veci,\vecj}$
 and $ \Delta^{\rm eff}_{\veci,\vecj}$ can then be considered as
  variational parameters
 which determine~$|\Psi_0\rangle$.  Note that, in the main part of this work, 
we will consider
 superconducting ground states with $d$-wave symmetry for which
 the local pairing amplitude vanishes, 
 \begin{equation}\label{456}
\langle \hat{c}^{(\dagger)}_{\veci,\uparrow}  
\hat{c}^{(\dagger)}_{\veci,\downarrow}  \rangle_{0}=
\langle \hat{c}^{(\dagger)}_{\veci,\uparrow}  
\hat{c}^{(\dagger)}_{\veci,\downarrow}  \rangle_{\rm G}=
0\;. 
\end{equation}
Here we introduced the notation $\langle \ldots \rangle_{0,{\rm G}}$  
  for expectation values with respect 
 to $|\Psi_0\rangle $ and $|\Psi_{\rm G}\rangle$. The case of a finite
 local pairing~(\ref{456}) is discussed in Appendix~\ref{app0}.
 
\subsection{Diagrammatic expansion}\label{yue}
We need to evaluate the expectation value
 of the Hamiltonian~(\ref{eq:1}),
\begin{equation}
E_{\rm G}
\equiv\sum_{\veci,\vecj,\sigma}t_{\veci,\vecj}
\frac{\langle \Psi_{\rm G} |\hat{c}_{\veci,\sigma}^{\dagger}
\hat{c}_{\vecj,\sigma}^{\phantom{\dagger}} | \Psi_{\rm G}  \rangle}
{\langle \Psi_{\rm G} | \Psi_{\rm G}  \rangle}
+ U\sum_{\veci}\frac{\langle \Psi_{\rm G} |\hat{d}_{\veci} | \Psi_{\rm G}  \rangle}
{\langle \Psi_{\rm G} | \Psi_{\rm G}  \rangle}
 \; ,
\label{eq:1.690}
\end{equation}
 with respect to our  Gutzwiller wave 
 function~(\ref{eq:1.2}). 
 As first shown in Ref.~\cite{buenemann2012a}, we can develop an 
 efficient diagrammatic scheme for this evaluation if we demand that 
 \begin{equation}
\hat{P}_{\vecl}^{\dagger} \hat{P}_{\vecl}^{}
=\hat{P}^2_{\vecl}=1+x\hat{d}_{\vecl}^{\rm HF} \; ,
\label{eq:1.6}
\end{equation}
where 
\begin{equation}\label{4565}
\hat{d}_{\vecl}^{\rm HF}\equiv 
\hat{n}^{\rm HF}_{\vecl,\uparrow}\hat{n}^{\rm HF}_{\vecl,\downarrow} \;\;,\;\;
\hat{n}^{\rm HF}_{\vecl,\sigma}\equiv\hat{n}_{\vecl,\sigma}-n_0\;,
\end{equation}
and $n_0\equiv\langle\hat{n}_{\vecl,\sigma}\rangle_0=N/(2L)$.
Equation~(\ref{eq:1.6}) determines three of the four 
parameters $\lambda_{\Gamma}$ as well as the coefficient~$x$.
In this way, we are left with only one variational parameter. 
For instance, we may express the parameters $\lambda_{\Gamma}$ 
by the coefficient 
 $x$, 
\begin{eqnarray}\label{dft1}
\lambda^2_d&=&1+x(1-n_0)^2 \;,\\\label{dft2}
\lambda^2_{\sigma}&=&1-xn_0(1-n_0) \;,\\
\lambda^2_{\emptyset}&=&1+xn_0^2 \;.
\end{eqnarray}

For the calculation of~(\ref{eq:1.690})  we need to evaluate
 three power series in $x$,
 {\arraycolsep=2pt\begin{eqnarray}
\langle\Psi_{\rm G}|\Psi_{\rm G} \rangle 
&=&\sum_{k=0}^{\infty}\frac{x^k}{k!}
\sideset{}{'}\sum_{\vecl_1,\ldots \vecl_k}
\bigl\langle \hat{d}^{\rm HF}_{\vecl_1,\ldots,\vecl_k}
\bigr\rangle_{0}
\label{eq:1.9}\, ,\\
\langle\Psi_{\rm G}|\hat{d}_{\veci}^{\vphantom{\rm HF}}|\Psi_{\rm G} \rangle 
&=&\lambda_d^2\sum_{k=0}^{\infty}\frac{x^k}{k!}
\sideset{}{'}\sum_{\vecl_1,\ldots \vecl_k}
\bigl\langle \hat{d}_{\veci}^{\vphantom{\rm HF}}\hat{d}^{\rm HF}_{\vecl_1,\ldots,\vecl_k}
\bigr\rangle_{0}
\label{eq:1.9b}\, ,
\\
\langle\Psi_{\rm G}|\hat{c}^{\dagger}_{\veci,\sigma}\hat{c}_{\vecj,\sigma}^{\phantom{\dagger}} 
|\Psi_{\rm G} \rangle&=&
\sum_{k=0}^{\infty}\frac{x^k}{k!}
\sideset{}{'}\sum_{\vecl_1,\ldots \vecl_k}
\bigl\langle
\widetilde{c}_{\veci,\sigma}^{\dagger}
\widetilde{c}_{\vecj,\sigma}^{\phantom{\dagger}}
\hat{d}^{\rm HF}_{\vecl_1,\ldots,\vecl_k}\bigr\rangle_{0} \, ,
\label{eq:1.9c}
\end{eqnarray}}
where we used Eq.~(\ref{eq:1.6}) and introduced the notation
\begin{eqnarray}
\hat{d}^{\rm HF}_{\vecl_1,\ldots,\vecl_k}&\equiv&\hat{d}^{\rm HF}_{\vecl_1}\cdots
\hat{d}^{\rm HF}_{\vecl_k}\quad, \quad
\hat{d}^{\rm HF}_{\emptyset}\equiv 1\;,\\\label{zxc}
\widetilde{c}_{\veci,\sigma}^{(\dagger)}&\equiv&
\hat{P}_{\veci}\hat{c}^{(\dagger)}_{\veci,\sigma}\hat{P}_{\veci}\;.
\end{eqnarray}
The primes in Eqs.~(\ref{eq:1.9})--(\ref{eq:1.9c}) indicate the summation 
restrictions 
\begin{equation}\label{sre}
\vecl_p\neq \vecl_{p'}\;, \; \;\vecl_p \neq  \veci, \vecj\; \; \;
\; \; \;\forall p,p'\;. 
\end{equation}

The expectation values $\langle \ldots  \rangle_0$ 
in~(\ref{eq:1.9})-(\ref{eq:1.9c})  can be evaluated by means of 
Wick's theorem~\cite{fetter2003}. In the resulting diagrammatic 
expansion,
the $k$th-order terms correspond to diagrams 
with $k$ `internal' vertices on sites $\vecl_1,\ldots,\vecl_k$, one (two) 
`external' vertices on site $\veci$ ($\veci$ and $\vecj$) 
and lines 
\begin{eqnarray}\label{pl}
P^{\sigma}_{\vecl,\vecl'}&\equiv& 
\langle\hat{c}^{\dagger}_{\vecl,\sigma}\hat{c}_{\vecl',\sigma}^{\phantom{\dagger}}\rangle_0
\;,\\
S_{\vecl,\vecl'}&\equiv& \label{pl5}
\langle\hat{c}^{\dagger}_{\vecl,\uparrow}
\hat{c}^{\dagger}_{\vecl',\downarrow}\rangle_0=\langle
\hat{c}_{\vecl',\downarrow}\hat{c}_{\vecl,\uparrow}\rangle^*_0
\end{eqnarray}
 connecting these vertices. By construction, 
we eliminated all diagrams with local `Hartree bubbles' 
at internal vertices, i.e., diagrams with lines that leave and enter 
 the same internal vertex.  To achieve the same 
for the external vertices in~(\ref{eq:1.9b}),(\ref{eq:1.9c}) 
 we rewrite the corresponding operators  as
\begin{eqnarray}
\hat{d}_{\veci}&=&(1-xd_0)\hat{d}^{\rm HF}_{\veci}
+n_0(\hat{n}^{\rm HF}_{\veci,\uparrow}+\hat{n}^{\rm HF}_{\veci,\downarrow}) 
+d_0\hat{P}^2_{\veci}
\label{eq:1.10a}\; ,
\\
\widetilde{c}_{\veci,\sigma}^{(\dagger)}&=&
q \hat{c}^{(\dagger)}_{\veci,\sigma}
+\alpha\hat{c}^{(\dagger)}_{\veci,\sigma}\hat{n}^{\rm HF}_{\veci,\bar{\sigma}}
\label{eq:1.10b}\; ,
\end{eqnarray}
where we introduced
\begin{eqnarray}
d_0&\equiv& n_0^2\;,\\
q&\equiv&\lambda_1(\lambda_dn_0+\lambda_{\emptyset}(1-n_0))\;,\\
\alpha&\equiv&\lambda_1(\lambda_d-\lambda_\emptyset)\;,
\end{eqnarray}
 and 
$\bar{\uparrow}=\downarrow$, $\bar{\downarrow}=\uparrow$.
When inserted into~(\ref{eq:1.9b}), the last term in~(\ref{eq:1.10a}) 
combines to $\lambda_d^2 d_0\langle \Psi_{\rm G}|\Psi_{\rm G}\rangle$
so that it does not have to be evaluated diagrammatically.

As a result, we obtain diagrammatic sums with no Hartree bubbles at {\sl any}
 vertex. This allows us to replace the lines~(\ref{pl}) by
   \begin{equation}\label{pl56}
\bar{P}^{\sigma}_{\vecl,\vecl'}\equiv 
\langle\hat{c}^{\dagger}_{\vecl,\sigma}\hat{c}_{\vecl',\sigma}^{\phantom{\dagger}}\rangle_0
-\delta_{\vecl,\vecl'}n_0\;.
\end{equation}
As demonstrated in Ref.~\cite{buenemann2012a}, 
the elimination of Hartree bubbles 
 has significant consequences for the convergence and, hence, accuracy of 
 our diagrammatic expansion. Due to our assumption of  
 d-wave superconductivity, we 
do not have to eliminate `anomalous' Hartree bubbles of the form~(\ref{456}).  
In Appendix~\ref{app0} we explain how our diagrammatic method can be generalised
 if Eq.~(\ref{456}) is not fulfilled.   

As the final analytical step of our derivation,  we apply 
the linked-cluster theorem~\cite{fetter2003}.
The norm~(\ref{eq:1.9}) cancels the disconnected diagrams 
in the two numerators~(\ref{eq:1.9b}) and~(\ref{eq:1.9c}). 
Note that for the application of this theorem, we first need to 
lift the summation restrictions in Eqs.~(\ref{eq:1.9})-(\ref{eq:1.9c}). 
This can be done, however, without generating additional terms, as we
 explain in Appendix~\ref{app1}.

For a translationally invariant system, the remaining task is to evaluate 
the diagrammatic sums
\begin{equation}\label{diagsss}
S=\sum_{k=0}^{\infty}\frac{x^k}{k!}S(k)
\end{equation}
with
\begin{equation}\label{diagss}
S\in\Big\{I^{(2)}, I^{(4)}, T^{(1),(1)}_{\veci,\vecj}, T^{(1),(3)}_{\veci,\vecj},
 T^{(3),(1)}_{\veci,\vecj}, T^{(3),(3)}_{\veci,\vecj}\Big \}
\end{equation}
and
\begin{eqnarray}\label{sd}
&&I^{(2)[(4)]}(k)\equiv\sum_{\vecl_1,\ldots, \vecl_k}
\bigl\langle 
\hat{n}^{\rm HF}_{\veci,\sigma}[\hat{d}^{\rm HF}_{\veci}]
\hat{d}^{\rm HF}_{\vecl_1,\ldots,\vecl_k}
\bigr\rangle^{\rm c}_{0}\; ,\\\nonumber
&&T_{\veci,\vecj}^{(1)[(3)],(1)[(3)]}(k)\\\label{sd2}
&&\equiv
\sum_{\vecl_1,\ldots,\vecl_k}
\bigl\langle
[\hat{n}^{\rm HF}_{\veci,\bar{\sigma}}]
\hat{c}^{\dagger}_{\veci,\sigma}
[\hat{n}^{\rm HF}_{\vecj,\bar{\sigma}}]
\hat{c}_{\vecj,\sigma}^{\phantom{\dagger}}\hat{d}^{\rm HF}_{\vecl_1,\ldots,\vecl_k}
\rangle^{\rm c}_{0}\;.
\end{eqnarray}%
Here, $\langle \dots \rangle_0^{ \rm c}$ indicates that only connected diagrams 
are to be kept. Note that in the evaluation of these diagrams, i.e., 
 after the application of the link-cluster theorem, one must not use 
 any summation restrictions as in~(\ref{eq:1.9b}) and~(\ref{eq:1.9c}), 
  see Appendix~\ref{app1}.   
 
The structure of the 
variational ground-state energy functional is the same 
 as in the paramagnetic case~\cite{buenemann2012a} and given by
\begin{equation}\label{355}
\langle \hat{H} \rangle_{\rm G}=E_{\rm G}(|\Psi_0\rangle,x)\equiv L(E^{\rm kin}+Ud)
\end{equation}
where
\begin{eqnarray}\label{356}
E_{\rm G}
&=& 2\sum_{\veci,\vecj}t_{\veci,\vecj}
\big(q^2T_{\veci,\vecj}^{(1),(1)}
+2q\alpha T_{\veci,\vecj}^{(1),(3)}
+\alpha^2 T_{\veci,\vecj}^{(3),(3)}\big)
\nonumber\\
&&+LU \lambda_d^2\big((1-xd_0)I^{(4)}+2n_0I^{(2)}+d_0\big) \; .
\label{eq:erz}
\end{eqnarray}
 This energy has to be minimised with respect 
 to $|\Psi_0\rangle $ and~$x$ where 
 $|\Psi_0\rangle $ enters the 
energy expression solely through the lines~(\ref{pl5}), (\ref{pl56})
 and through $n_0$. Note that in the presence of superconductivity,
  the particle number
 per lattice site
\begin{eqnarray}\label{2sep}
n_{\rm G}&\equiv&\langle \hat{n}_{\veci,\sigma} \rangle_{\rm G}=
\lambda^2_{d}\big(d_0+I^{(4)}(1-xd_0)+2n_0I^{(2)}\big)\\\nonumber
&&+\lambda^2_{1}
\big(m^0_{1}+I^{(2)}(1-2n_0)-I^{(4)}(1+xm^0_{1})\big)\;,
\end{eqnarray}
with $m^0_{1}=n_0(1-n_0)$ is not the same as 
$n_{0}$. Physically, however, 
 the value of $n_{\rm G}$ rather than $n_0$ should be fixed in 
the minimisation 
 of the energy. Therefore, we  minimise the grand-canonical potential
  \begin{equation}
 \mathcal{F} =
E_{\rm G} - 2 \mu_{\rm G} n_{\rm G} L
\end{equation}
with respect to $\bar{P}^{\sigma}_{\vecl,\vecl'}$, $ S_{\vecl,\vecl'}$, $n_0$, and 
 $x$ where the chemical potential 
 $\mu_{\rm G}$ allows us to vary the correlated particle number.  
  The minimisation with respect to 
$\bar{P}^{\sigma}_{\vecl,\vecl'}$, $ S_{\vecl,\vecl'}$, and $n_0$ leads to the 
 effective single-particle equation for $|\Psi_0\rangle$,
  \begin{equation}\label{rde}
\hat{H}_0^{\rm eff}|\Psi_0\rangle=E_0  |\Psi_0\rangle
\end{equation}
 with a Hamiltonian $\hat{H}_0^{\rm eff}$ 
as introduced in~(\ref{eq:iou1}) and parameters  
\begin{eqnarray}\label{eq:iouFF}
t^{\rm eff}_{\veci,\vecj} &=&
\frac{\partial \mathcal{F} (|\Psi_0\rangle,x)}{\partial P_{\veci,\vecj}}
\;\;\;({\rm for}\;\; \veci \neq \vecj)\;,\\
t^{\rm eff}_{\veci,\veci} &=&
\frac{\partial \mathcal{F} (|\Psi_0\rangle,x)}{\partial n_0}\;,
\\
 \quad \Delta^{\rm eff}_{\veci,\vecj} &=&
\frac{\partial \mathcal{F} (|\Psi_0\rangle,x)}{\partial S_{\veci,\vecj}} \;.
\label{eq:iouF}
\end{eqnarray}

Equations~(\ref{356})-(\ref{eq:iouF}) need to be solved 
 self-consistently, together with the minimisation condition
\begin{equation}
\frac{\partial}{\partial x}\mathcal{F} (|\Psi_0\rangle,x)=0.
\end{equation}
Numerically, this can been achieved by the following iterative procedure:
\begin{itemize}
\item[(i) ] Chose an initial value for 
$|\Psi^{\rm i}_0\rangle \equiv |\Psi_0\rangle$.
\item[(ii) ] Determine the variational parameter $x_{\rm min}$ which 
 minimises $E_{\rm G}$ for a fixed $|\Psi_0\rangle=|\Psi^{\rm i}_0\rangle$.
\item[(iii) ] Determine the parameters~(\ref{eq:iouFF})-(\ref{eq:iouF}) and the
 corresponding Hamiltonian $\hat{H}_0^{\rm eff}$  
 for $x=x_{\rm min}$. 
\item[(iv) ] Determine the ground state  $|\Psi^{\rm f}_0\rangle$ of 
$\hat{H}_0^{\rm eff}$\;.
\item[(v) ] If $|\Psi^{\rm f}_0\rangle \approx |\Psi^{\rm i}_0\rangle$ terminate 
 the algorithm. Otherwise, set 
$|\Psi^{\rm i}_0\rangle = |\Psi^{\rm f}_0\rangle$ and go back to
 point (ii).
\end{itemize} 
Note that, numerically, it is usually  necessary  to introduce 
 some form of `damping' in the calculation of the 
Hamiltonian $\hat{H}_0^{\rm eff}$
  in step iii):  If $\hat{H}_{a}^{\rm eff}$ has been used 
 in the previous iteration and 
  $\hat{H}_{b}^{\rm eff}$ is the Hamiltonian derived
 from~(\ref{eq:iouFF})-(\ref{eq:iouF}) 
 then one continues the procedure in iv) with 
\begin{equation}
\hat{H}_0^{\rm eff}\equiv \hat{H}_{a}^{\rm eff}+
\beta (\hat{H}_{b}^{\rm eff}-\hat{H}_{a}^{\rm eff})\;.
\end{equation}
Working with a parameter $\beta < 1$  ensures
 the convergence of our algorithm.

\subsection{Calculation of diagrams}\label{qwe}
 \begin{figure}[t] 
\includegraphics[width=8cm]{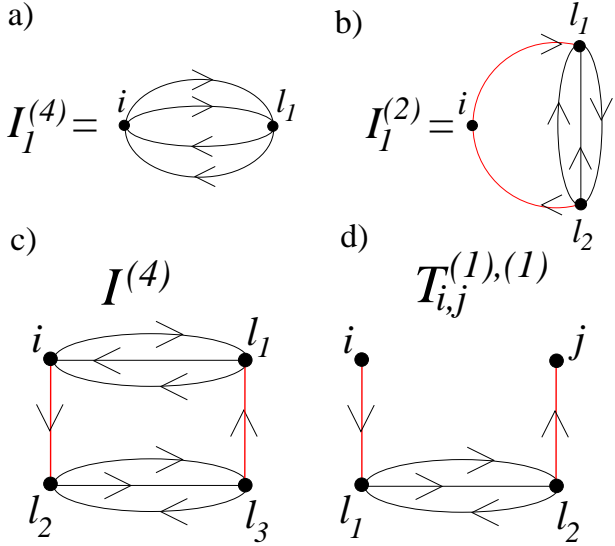} 
\caption{Lowest order diagram of a) $I^{(4)}$, b) $I^{(2)}$ and 
examples for long-range diagrams in c) $I^{(4)}$ and d) $T_{\veci,\vecj}^{(1),(1)}$.
\label{Fig:fig1}}
\end{figure} 
To carry out the minimisation, as described in the previous 
 section, we need to calculate the diagrams~(\ref{diagss}) 
 and their derivatives with respect to lines
 up to a certain order in $x$. For example, the first-order
 diagram  $I^{(4)}_1$ of $I^{(4)}$ is shown in Fig.~\ref{Fig:fig1}a). 
Here, the lines can be normal ($P^{\sigma}_{\veci,\vecl_1}$) or 
 anomalous ($S_{\veci,\vecl_1}$). With four normal lines, e.g.,
 we have to evaluate
\begin{eqnarray}
I^{(4)}_1&=&\sum_{\vecl_1(\ne \veci)}P^{\uparrow}_{\veci,\vecl_1}P^{\uparrow}_{\vecl_1,\veci}
P^{\downarrow}_{\veci,\vecl_1}P^{\downarrow}_{\vecl_1,\veci}\\
&=&\frac{1}{L^3}\sum_{\veck,\veck',\veck^{\prime \prime} }
 n_{\veck,\uparrow} n_{\veck',\uparrow} 
 n_{\veck^{\prime \prime} ,\downarrow}n_{\veck+\veck'+\veck^{\prime \prime} ,\downarrow}-n_0^4 
\end{eqnarray}
where we have introduced the momentum-space distribution
  $n_{\veck,\sigma}\equiv \langle  
\hat{c}^{\dagger}_{\veck,\sigma}   \hat{c}^{}_{\veck,\sigma}  \rangle_0$.
Obviously, the real space evaluation of the diagram is numerically much 
easier because only one summation 
(over $\vecl_1$) has to be carried out. Moreover, 
the lines $P_{\veci,\vecj}$ 
in real space vanish like $1/\sqrt{|\veci-\vecj|}$ while the number 
  of neighbours of this distance is $\sim |\veci-\vecj|$. Hence, the 
 real-space contributions of $I^{(4)}_1$ fall off rapidly and we can restrict 
 the summation over $l_1$ to a limited number of nearest neighbours of $\veci$. 
 This `locality' of diagrams in real space is a key ingredient 
  in our numerical implementation since it allows us to calculate 
 diagrams up to relatively large orders in $x$. 

Unfortunately, not all diagrams 
 are as local as $I^{(4)}_1$. In particular, all diagrams in $I^{(2)}$, contain 
 `long-range contributions', e.g., the joint sum over 
$\vecl_1$, $\vecl_2$ in  Fig.~\ref{Fig:fig1}b).  
In the paramagnetic case, there exists a relationship between $I^{(2)}$ and 
 $I^{(4)}$ which can be used to circumvent the long-range contributions 
 in  $I^{(2)}$, see  Ref.~\cite{buenemann2012a}. 
No such relationship, however, can be used 
 for superconducting states. Moreover, even for paramagnetic states some diagrams
  have long-range contributions, e.g., the $I^{(4)}$ and $T_{\veci,\vecj}^{(1),(1)}$ 
 diagrams shown in Fig.~\ref{Fig:fig1}. 
We may identify the diagrams with long-range contributions  by a 
topological analysis, as we shall explain in the following Section~\ref{app4}. The 
real-space summations  which belong to such long-range diagrams can then be 
 evaluated analytically, see below.


\section{Long-range diagrams}\label{app4}
\begin{figure}[t] 
\includegraphics[width=8cm]{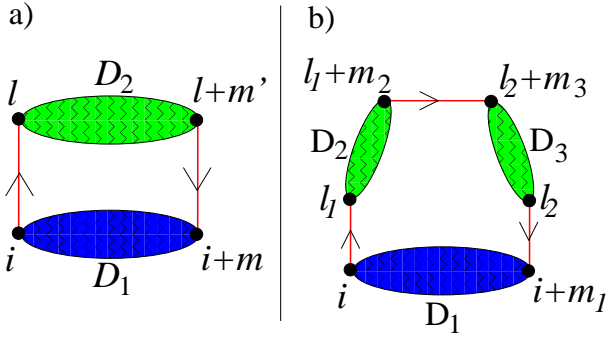} 
\caption{Long-range diagrams of a) type I, b) type II.
\label{Fig:fig11}}
\end{figure}

\begin{figure}[t] 
\includegraphics[width=5cm]{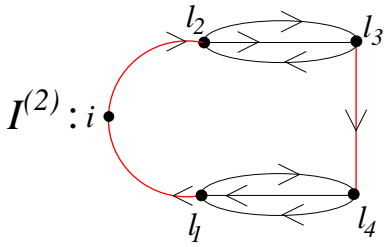} 
\caption{A type II diagram in $I^{(2)}$.
\label{Fig:fig12}}
\end{figure}

As explained in Section~\ref{qwe}, some diagrams are not 
  localised and require a special treatment in our real-space evaluation.
 Topologically, there are
two types of diagrams which we need to consider
 up to the 4-th order (of internal vertices) in $x$. They are displayed in 
 Figs.~\ref{Fig:fig11}.  
Diagrams of type I can be split into two disconnected diagrams $D_1$ and $D_2$ by 
cutting two lines,
where the external vertices ($\veci,\vecj$) belong to $D_1$. 
 Examples for such diagrams are shown in  Figs.~\ref{Fig:fig1}b)-d).
 In a similar way we define diagrams of type II as those which can be  
 split into three disconnected diagrams by cutting three (single) lines. An example 
 for this type is the $I^{(2)}$ diagram shown in Fig.~\ref{Fig:fig12}. It illustrates that 
 type II diagram can only appear if there are at least four internal vertices. 
 Note that more complicated long-range diagrams 
require the inclusion of more 
 than five internal vertices.  
 
For the evaluation of the long-range diagrams in Figs.~\ref{Fig:fig11}, 
one can carry out the sums 
 over $\vecl$ (type I) or $\vecl_1, \vecl_2$ (type II)  analytically. We will consider 
 the paramagnetic and the superconducting case separately in the 
 following two sections.

 \subsection{The paramagnetic case}\label{app4a}
\begin{figure}[t] 
\includegraphics[width=6cm]{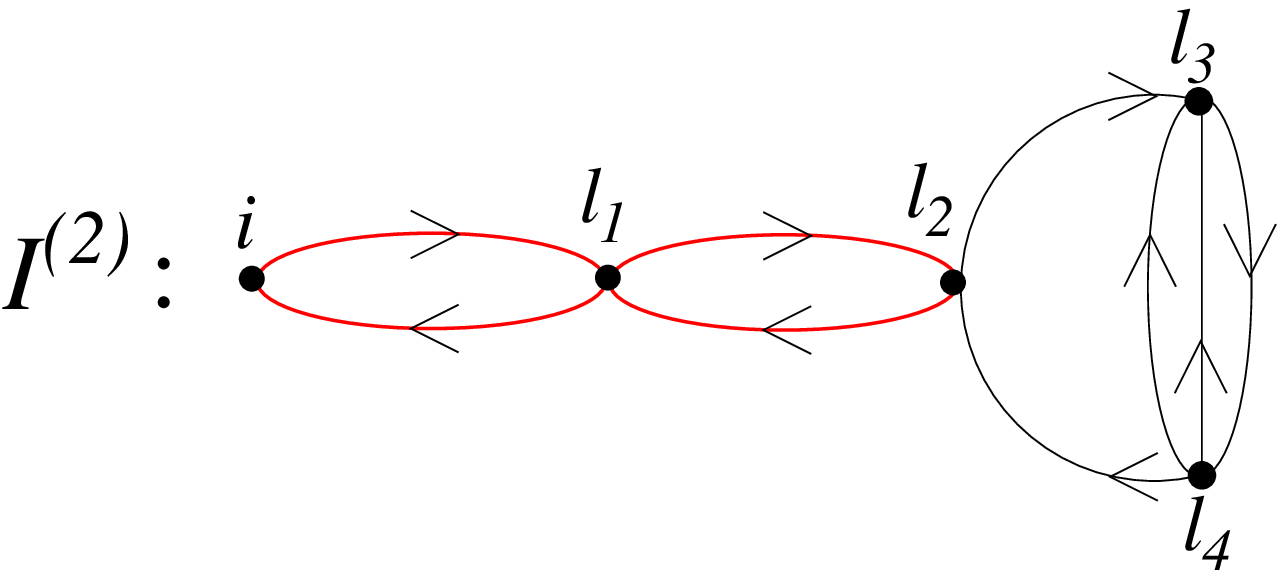} 
\caption{A type I diagram in $I^{(2)}$.
\label{Fig:fig13}}
\end{figure}

For the evaluation of the diagram in Fig.~\ref{Fig:fig11}a), we need to calculate
\begin{equation}
  D_{I}=\sum_{\vecm,\vecm'} D^1_{\vecm,\veco}D^2_{\veco,\vecm'}\sum_{\vecl}
\bar{P}^{\sigma}_{\veco,\vecl}\bar{P}^{\sigma}_{\vecl+\vecm',\vecm}
\end{equation}
where
  $D^1$, $D^2$ are assumed to be localised diagrams, i.e., the sums over $\vecm,\vecm'$ 
 can be restricted to a shell around~$\veco$.  
 The sum over $\vecl$ yields
\begin{eqnarray}\nonumber 
&&\bar{D}(\vecm,\vecm')\equiv 
\sum_{\vecl}
\bar{P}^{\sigma}_{\veco,\vecl}\bar{P}^{\sigma}_{\vecl+\vecm',\vecm}\\\nonumber
&&= \sum_{\vecl}
(P^{\sigma}_{\veci,\vecl}-\delta_{\veci,\vecl}n_0)
(P^{\sigma}_{\vecl+\vecm',\veci+\vecm}-\delta_{\vecl+\vecm',\veci+\vecm}n_0)\\\label{qxr}
&&=(1-2n_0)P^{\sigma}_{\vecm,\vecm'}+\delta_{\vecm,\vecm'}n_0^2 \;.
\end{eqnarray}
Here we have used
\begin{equation}
\sum_{\vecl}P^{\sigma}_{\veci,\vecl}P^{\sigma}_{\vecl+\vecm',\veci+\vecm}=P^{\sigma}_{\vecm,\vecm'}\;,
\end{equation}
which holds because, after Fourier transformation, in  the 
 paramagnetic case we can use $n_{\veck,\sigma}^2=n_{\veck,\sigma}$ 
in momentum-space. 
A long range-diagram of type I is therefore given as 
\begin{equation}\label{qxr2}
  D_{I}=\sum_{\vecm,\vecm'} D^1_{\vecm,\veco}D^2_{\veco,\vecm'}\bar{D}(\vecm,\vecm') \;.
\end{equation}
Note that the diagram  $D^1$ may contain additional long-range elements as, e.g., in the 
 $I^{(2)}$ diagram in Fig.~\ref{Fig:fig13}. In such a case, Eqs.~(\ref{qxr})-(\ref{qxr2})
 have to be applied consecutively.  

In a long-range diagram of type II we need to evaluate
\begin{equation}
   D_{II}=\sum_{\vecm_1,\vecm_2,\vecm_3} D^1_{\vecm_1,\veco}D^2_{\veco,\vecm_2}
D^3_{\vecm_3,\veco}\bar{D}(\vecm_1,\vecm_2,\vecm_3)
\end{equation}
where
\begin{equation}\label{dfgh2}
\bar{D}(\vecm_1,\vecm_2,\vecm_3)\equiv \sum_{\vecl_1,\vecl_2}
\bar{P}^{\sigma}_{\veco,\vecl_1}\bar{P^{\sigma}}_{\vecl_1+\vecm_2,\vecl_2+\vecm_3}\bar{P}^{\sigma}_{\vecl_2,\vecm_1}\;,
\end{equation}
The sums over $\vecl_1$,  $\vecl_2$ can again be calculated exactly. This leads to 
\begin{eqnarray}\nonumber
\bar{D}(\vecm_1,\vecm_2,\vecm_3)&=&(1-3n_0+3n_0^2)P^{\sigma}_{m_2,m_1+m_3}\\
&&-\delta_{m_2,m_1+m_3}n_0^3\;.
\end{eqnarray}

 \subsection{The superconducting case}\label{app4b}
In the superconducting case, the single (red) lines in Figs.~\ref{Fig:fig11} 
 can be normal or anomalous. We therefore introduce the 
 abbreviations
\begin{eqnarray}\label{34r1}
X^1_{\veci,\vecj}&\equiv& P^{\sigma}_{\veci,\vecj}\;,\\\label{34r2}
X^2_{\veci,\vecj}&\equiv& S_{\veci,\vecj}\;,\\
\bar{X}^{\alpha}_{\veci,\vecj}&\equiv& X^{\alpha}_{\veci,\vecj}-\delta_{\veci,\vecj}X^{\alpha}_{\veco,\veco}\;,
\end{eqnarray}
and the corresponding Fourier transforms
\begin{eqnarray}
X^1_{\veck}&\equiv& n_{\veck,\sigma}\;,\\
X^2_{\veck}&\equiv& \langle \hat{c}_{\veck,\uparrow} \hat{c}_{-\veck,\downarrow}    \rangle_{0}\;.
\end{eqnarray}
For long-range diagrams of type I we then have to evaluate 
 \begin{eqnarray}
  D^{\alpha,\alpha'}_{I}&=&\sum_{\vecm,\vecm'} D^1_{\vecm,\veco}D^2_{\veco,\vecm'}
\bar{D}^{\alpha,\alpha'}(\vecm,\vecm')\\
\bar{D}^{\alpha,\alpha'}(\vecm,\vecm')&\equiv&\sum_{\vecl}
\bar{X}^{\alpha}_{\veco,\vecl}\bar{X}^{\alpha'}_{\vecl+\vecm',\vecm}\;,
\end{eqnarray}
where $\alpha^{(\prime)}\in \{1,2\}$ characterises the 
 two single lines in Fig.~\ref{Fig:fig11}a). Note that 
 $X^2_{\veco,\veco}=0$ for our d-wave states. 
With a transformation to momentum space  we find
 \begin{eqnarray}\nonumber
\bar{D}^{\alpha,\alpha'}(\vecm,\vecm')&=&Y^{\alpha,\alpha'}_{\vecm,\vecm'}
-X^{\alpha}_{\veco,\veco}X^{\alpha'}_{\vecm,\vecm'}
-X^{\alpha'}_{\veco,\veco}X^{\alpha}_{\vecm,\vecm'}\\\label{38r2}
&&+\delta_{\vecm,\vecm'} X^{\alpha}_{\veco,\veco}X^{\alpha'}_{\veco,\veco}
\end{eqnarray}
with
\begin{equation}
Y^{\alpha,\alpha'}_{\vecm,\vecm'}\equiv\frac{1}{L} \sum_{\veck}
X^{\alpha}_{\veck}X^{\alpha'}_{\veck}e^{{\rm i}\veck(\vecm-\vecm')}\;.
\end{equation}
Note that, in the paramagnetic case, we have
$Y^{1,1}_{\vecm,\vecm'}=P^{\sigma}_{\vecm,\vecm'}$, $X^{1}_{\veco,\veco}=n_0$, 
 such that Eq.~(\ref{qxr}) is recovered. 

The evaluation of type II diagrams leads to 
\begin{eqnarray}\nonumber
 D^{\alpha_1,\alpha_2,\alpha_3}_{II}&=&\sum_{\vecm_1,\vecm_2,\vecm_3} D^1_{\vecm_1,\veco}D^2_{\veco,\vecm_2}
D^3_{\vecm_3,\veco}\\
&&\times\bar{D}^{\alpha_1,\alpha_2,\alpha_3}(\vecm_1,\vecm_2,\vecm_3)
\end{eqnarray}
where
\begin{eqnarray}\nonumber
&&\bar{D}^{\alpha_1,\alpha_2,\alpha_3}(\vecm_1,\vecm_2,\vecm_3)\\\label{345}
&&\equiv \sum_{\vecl_1,\vecl_2}
\bar{X}^{\alpha_1}_{\veco,\vecl_1}\bar{X}^{\alpha_2}_{\vecl_1+\vecm_2,\vecl_2+\vecm_3}
\bar{X}^{\alpha_3}_{\vecl_2,\vecm_1}\;.
\end{eqnarray}
The momentum-space evaluation for~(\ref{345}) yields
\begin{eqnarray}\nonumber
&&\bar{D}^{\alpha_1,\alpha_2,\alpha_3}(\vecm_1,\vecm_2,\vecm_3)=
Z^{\alpha_1,\alpha_2,\alpha_3}_{m_2,m_1+m_3}\\\nonumber
&&-[X^{\alpha_1}_{\veco,\veco}Y^{\alpha_2,\alpha_2}_{m_2,m_1+m_3}+{\rm perm.}]\\\nonumber
&&+[X^{\alpha_1}_{\veco,\veco}X^{\alpha_2}_{\veco,\veco}X^{\alpha_3}_{m_2,m_1+m_3}+{\rm perm.}]\\\label{3458}
&&-\delta_{\vecm_2,\vecm_1+\vecm_3}X^{\alpha_1}_{\veco,\veco}X^{\alpha_2}_{\veco,\veco}X^{\alpha_3}_{\veco,\veco}
\end{eqnarray}
where `$+$perm.' denotes the three cyclical permutations of $(\alpha_1,\alpha_2,\alpha_3)$
 in the respective functions and 
 \begin{equation}
Z^{\alpha_1,\alpha_2,\alpha_3}_{\vecm,\vecm'}\equiv\frac{1}{L} \sum_{\veck}
X^{\alpha_1}_{\veck}X^{\alpha_2}_{\veck}X^{\alpha_3}_{\veck}e^{{\rm i}\veck(\vecm-\vecm')}\;.
\end{equation} 

Note that in the superconducting case it is not possible to write the long-range contributions 
 in terms of lines ($P^{\sigma}_{\veci,\vecj}$, $S_{\veci,\vecj}$) 
as it is possible in the paramagnetic case, see Eqs.~(\ref{qxr}) 
 and~(\ref{dfgh2}). Instead, there appear the new objects  
$Y^{\alpha,\alpha'}_{\vecm,\vecm'}$ and  
$Z^{\alpha_1,\alpha_2,\alpha_3}_{\vecm,\vecm'}$ whose calculation, however, is numerically 
 benign because $|\vecm-\vecm'|$ can be assumed to be small. Still, due to 
 the appearance of these new objects, we need to reconsider the minimisation 
 of our energy functional with respect to $|\Psi_0\rangle$. This problem 
is  discussed in Appendix~\ref{app5}.

 \begin{figure}[t] 
\includegraphics[width=6cm,angle=270]{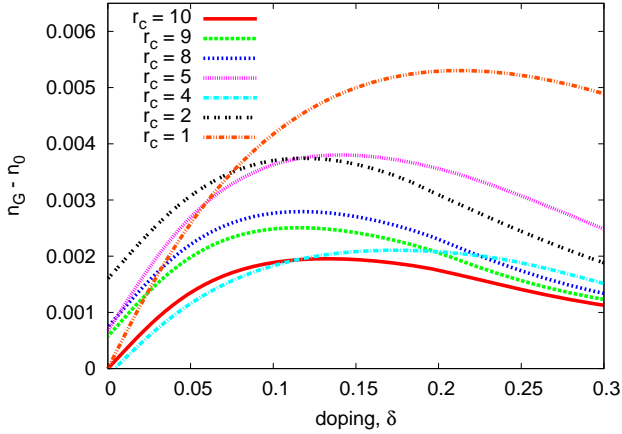} 
\caption{The $7$-th order Taylor 
 expansion of~(\ref{se6ad}) 
as a function of doping $\delta$
for $U/|t|=10$ and different values of the cutoff parameter $r_{\rm c}=1-10$ (calculated without LRDE).
\label{Fig:fig100}}
\end{figure} 

\section{Results}\label{se3}
In our real-space evaluation of diagrams there are two main 
 approximations that are needed  to make the problem  
 numerically treatable:
\begin{itemize}
\item[i) ] The lines $\bar{P}^{\sigma}_{\vecl,\vecl'}$, $ S_{\vecl,\vecl'}$  
which enter the energy functional can have an arbitrary 
`length' 
\[
|\vecl-\vecl'|=\sqrt{(l_1-l'_1)^2+(l_2-l'_2)^2}\;. 
\]
To keep the number of real-space contributions 
finite 
 we  need to introduce some `cutoff' $r_{\rm c}$, i.e., the assumption that  
$\bar{P}^{\sigma}_{\vecl,\vecl'}=S_{\vecl,\vecl'}=0$  for  $|\vecl-\vecl'|^2>r_{\rm c}$. 
The cutoff leads to numerical errors in particular for the long-range 
 diagrams discussed in  Section~\ref{app4}. However, as we will demonstrate in 
 the following section~\ref{se6a}, these errors are negligible if we 
 employ the long-range diagram evaluation (LRDE) technique
 introduced in Section~\ref{app4}.
\item[ii) ] The number of diagrams grows exponentially with the 
 number of internal vertices (index $k$ in Eqs.~(\ref{sd}), (\ref{sd2})). 
 Therefore, the diagrammatic expansion must be terminated at some finite 
 value of $k$. As we will discuss in  section~\ref{se6b}, a better truncation 
 parameter is the total number of lines in a diagram.  
 \end{itemize}  
In all subsequent results, we have worked with a single-band Hamiltonian 
$\hat{H}_0$
 that contains nearest and next-nearest neighbour hopping of 
 $t=-0.35$eV and $t'/t=-0.25$, respectively. These values are generally assumed 
 to describe  the situation in  the Cuprates.

\begin{figure}[t] 
\includegraphics[width=6cm,angle=270]{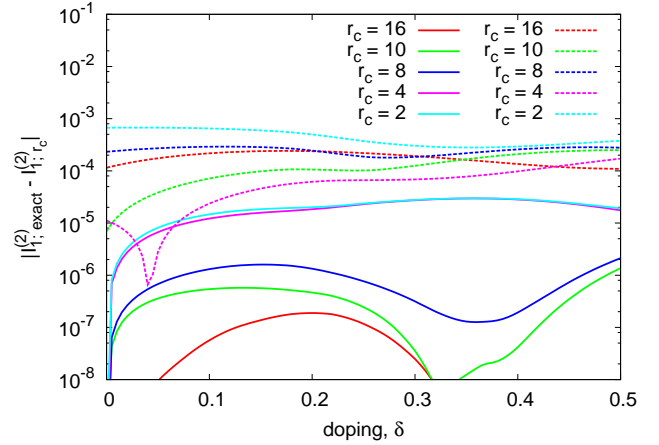} 
\caption{Difference between the exact value of the diagram 
$I^{(2)}_1$ (see Fig.~\ref{Fig:fig1}b))  as a function of 
doping $\delta$ for several values 
 of $r_{\rm c}$ with and without LRDE (solid and dashed lines, respectively).
\label{Fig:fig110}}
\end{figure}

 \subsection{Line cutoff}\label{se6a}
To analyse the role of a finite  cutoff length $r_{\rm c}$ 
 we first  consider the expression for the correlated particle number 
 per lattice site
\begin{equation}\label{se6ad}
n_{\rm G}-n_0=[1+xn_0(1-n_0)]I^{(2)}+x(1-2n_0)I^{(4)}
\end{equation}
which results from Eqs.~(\ref{dft1}), (\ref{dft2}),~(\ref{2sep}). 
The correlation operator $\hat{P}_{\rm G}$ in~(\ref{eq:1.2}) commutes with the
 operator 
 $\hat{N}\equiv \sum_{i,\sigma} \hat{n}_{i,\sigma}$ which counts the total 
 number of electrons. Since, in the paramagnetic case, $|\Psi_0 \rangle$ 
 is an eigenstate of $\hat{N}$, we have 
$\langle\hat{N} \rangle_{\rm G}= \langle\hat{N} \rangle_{0}$ and therefore 
  $n_{\rm G}-n_0=0$ in 
   our translationally invariant system.

If we consider the
 r.h.s. of~(\ref{se6ad}) as a power series in $x$ each coefficient 
 of this expansion has to be exactly zero. Numerically, however, this 
 is not the case because of our cutoff parameter $r_{\rm c}<\infty$. 
 In Fig.~\ref{Fig:fig100}, we plot the $7$-th order Taylor 
 expansion of~(\ref{se6ad}) as a function of doping 
for $U/|t|=10$ and different values of  $r_{\rm c}$.

As we can see from this figure, an increase of  $r_{\rm c}$
 improves the results only slowly. Unfortunately,  for higher-order 
 diagrams it would not be feasible to work with values of $r_{\rm c}$ significantly larger than 
 $10$. Therefore, the LRDE is essential to eliminate the error that stems from the finite
 cutoff $r_{\rm c}$. 
 In fact, the LRDE  ensures that the  $7$-th order expansion
 in  Fig.~\ref{Fig:fig100} is exactly zero for all cutoff parameters $r_{\rm c}$. 
  This perfect agreement, however, is only due to the fact that the remaining errors in the 
 calculation of $I^{(2)}$ and  $I^{(4)}$ cancel each other exactly in~(\ref{se6ad}).
 To gauge the remaining real-space cutoff error, 
we display in Fig.~\ref{Fig:fig110} the value of the 
diagram
 $I^{(2)}_1$ (see Fig.~\ref{Fig:fig1}b)) relative to its exact value as a function of 
doping for several values 
 of $r_{\rm c}$ with and without LRDE. Note that these values 
are independent of  $x$  (cf. Eq.~(\ref{sd})) and only depend on the lines which we take from 
 the ground state $|\Psi_0\rangle$ of the bare single particle Hamiltonian $\hat{H}_0$. 
Again, we can see from Fig.~\ref{Fig:fig110}
 that bringing down the numerical error by increasing $r_{\rm c}$ is not working well
 without the  use of the LRDE: even for $r_{\rm c}=2$ 
(i.e., only nearest and next-nearest neighbour lines) 
the results with LRDE are more 
 accurate than those for  $r_{\rm c}=16$ and without LRDE.  

  \begin{figure}[t] 
\includegraphics[width=5.7cm,angle=270]{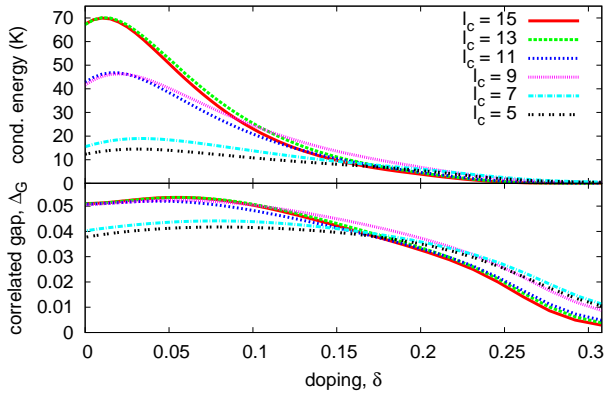} 
\caption{Condensation energy in Kelvin (upper panel)  and correlated gap 
in the superconducting state 
(lower panel) 
as a function of doping $\delta$ for $U/|t|=10$ and different values of $l_{c}$. 
\label{Fig:fig120}}
\end{figure}

\begin{figure}[t] 
\includegraphics[width=5.7cm,angle=270]{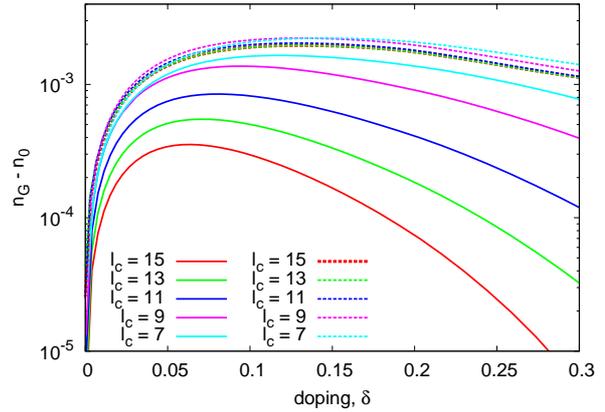} 
\caption{Difference $n_{\rm G}-n_0$ in the paramagnetic phase as a function of doping $\delta$ for $U/|t|=10$
 and several values of $l_{c}$, 
  with and without LRDE (solid and dashed lines, respectively).
\label{Fig:fig130}}
\end{figure} 

\subsection{Line number truncation}\label{se6b}
\begin{figure}[t] 
\includegraphics[width=6.0cm,angle=270]{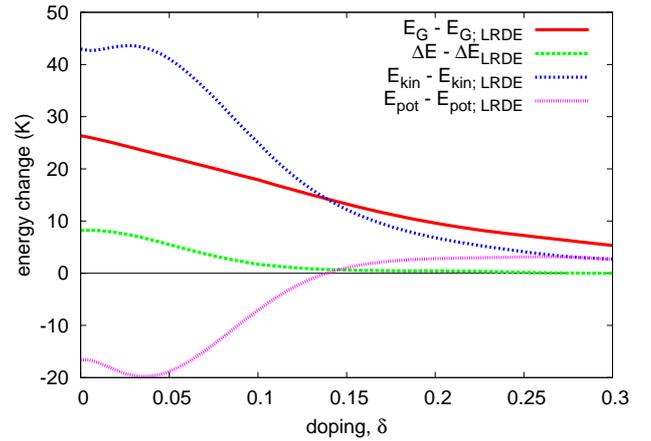} 
\caption{Differences between kinetic ($E_{\rm kin}$), potential ($E_{\rm pot}$), 
total ($E_{\rm G}$), and condensation ($\Delta E$)  energy 
calculated with and without the LRDE as a function of doping $\delta$ and for $U/|t|=10$.
\label{Fig:fig140}}
\end{figure}

The natural expansion parameter appears to be the number $k$ of internal vertices in 
 Eqs.~(\ref{diagsss})-(\ref{sd}). In fact, in our previous works on 
 Pomeranchuk phases~\cite{buenemann2012a} and superconductivity~\cite{buenemann2012b}   
we have investigated the convergence of results as a function of the diagrammatic order $k$. 
 However, the topological complexity of a diagram is more related to the number of 
 lines in a diagram which is  given as
\begin{eqnarray} \label{hfop77} 
N_k&=& 1+2k \;\;\;{\rm for}\;I^{(2)}[k],T^{(1)(1)}[k]\;,\\
N_k&=& 2+2k \;\;\;{\rm for}\;I^{(4)}[k],T^{(1)(3)}[k]\;,\\
N_k&=& 3+2k\;\;\;{\rm for}\;T^{(3)(3)}[k]\;.
\end{eqnarray}
As we have found already in a study on the $t$-$J$ model~\cite{kaczmarczyk2014a}  
it is more useful
 to include all diagrams up to a certain maximum number $l_{c}$ of lines. This means 
that in results with $l_{c}=15$ there are some diagrams included that have $k=7$
 internal vertices.  
As an example for the convergence with respect to $l_{c}$, we show in Fig.~\ref{Fig:fig120} 
the condensation energy (energy difference between superconducting and 
paramagnetic ground state) and the `correlated gap' 
 $\Delta_{\rm G}\equiv \langle \hat{c}_{\veci,\uparrow}  \hat{c}_{\vecj,\downarrow} \rangle_{\rm G}$ 
 (for nearest neighbours $\veci$, $\vecj$)
in the superconducting state for $U/|t|=10$ and 
as a function of doping for different values of $l_{c}$. All these data  have been 
 calculated using the LRDE (without LRDE we obtain qualitatively similar behavior, with the value of the condensation energy slightly increased, cf. also Fig. \ref{Fig:fig140}). As observed in previous studies, convergence of the condensation energy is reached for
 doping $\delta \gtrsim 0.1 $. For smaller doping values the convergence
 is less satisfactory as compared to that of most other observables, 
 e.g., the correlated gap. Figure~\ref{Fig:fig120} shows that the results for the latter have 
 converged already for $l_{c}=11$. Since the condensation energy  is  largely increasing as a function of $l_{c}$, the stability of a superconducting state 
 is very likely in  the 	inaccessible limit $l_{c}\to \infty$. Note that the ground state energy of 
the paramagnetic phase (not plotted) is practically converged for $l_{c}=15$ (the differences in this 
energy between the results for $l_{c}=15$ and $l_{c}=11, 13$ are below $1$K). Therefore, the error 
of the condensation energy comes mostly from the superconducting phase ground state energy.
 In the following analysis
 we  work with $l_{c}=15$ and $r_{c}=10$, unless stated otherwise.

The importance of the LRDE is illustrated, once more, in Fig.~\ref{Fig:fig130} where 
 we display $n_{\rm G}-n_0$ in the paramagnetic phase as a function of doping for $U/|t|=10$
 and several values of $l_{c}$. The solid (dashed) lines show the results  with (without)
 the LRDE. Obviously, the error that appears in the data without LRDE is so large, that 
 there is hardly any improvement of the results if we increase $l_{c}$. In contrast,  
  if we use the  LRDE,   $n_{\rm G}-n_0$ goes exponentially 
to zero if we increase  $l_{c}$.
 Note that the remaining error stems from the fact, that contributions from the 
 highest order in $x$ are not exactly cancelled as they were in the Taylor-series 
   expansion discussed in the previous section~\ref{se6a}.

In Fig.~\ref{Fig:fig140} we display the differences between energies that were calculated
 with and without the LRDE as a function of doping and for $U/|t|=10$.
 The differences shown in this figure are those for the kinetic, the potential and the 
 total energy in the superconducting phase,  as well as the condensation energy. 
As can be seen from this graph, the changes of the 
 energies due to the LRDE are not negligible, in particular, for 
 a doping of less than $0.1$. The total energy is lowered by up to
 $23$K if we use LRDE. The resulting change in the condensation energy, however, 
 is smaller (less than $10$K). The same holds for other observables in the superconducting state. 
 Therefore, the main results on superconductivity which we have published in our previous
 work~\cite{buenemann2012b}  remain unchanged by the new LRDE scheme. 
  However, the relatively large changes of the  kinetic and the potential energy indicate 
  that for other systems or states, the LRDE may alter physical properties more visibly.  

\begin{figure}[t!] 
\includegraphics[width=6.0cm,angle=270]{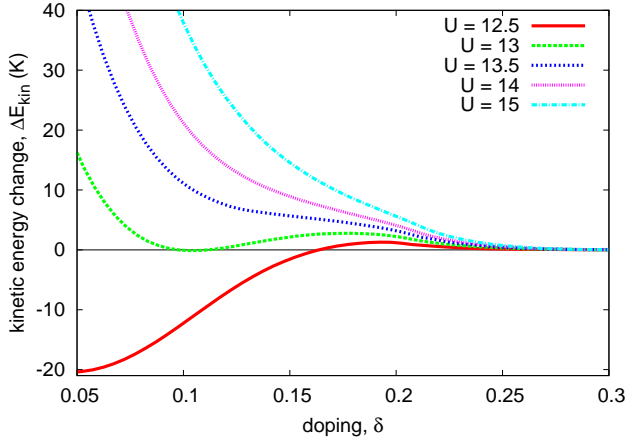} 
\caption{Kinetic energy change upon condensation $\Delta E_{\rm kin}$ as a function 
of doping $\delta$.
\label{Fig:fig150}}
\end{figure}

In Fig.~\ref{Fig:fig150} we show the difference between the kinetic energy of the paramagnetic and the superconducting phases. This property is negative for a conventional (BCS-type) superconductor, as pairing induces 'smearing' of the single-particle distribution around the Fermi surface, which increases the kinetic part of the total energy. We observe such behaviour for $U \lesssim 12 |t|$ at all doping values. For larger values of the Coulomb interaction ($U \gtrsim 13 |t|$) the kinetic energy becomes lower in the superconducting phase. This is an unconventional behaviour coming from the fact that the bandstructure of the effective Hamiltonian changes upon condensation, which can dominate over the mentioned effect of an increased kinetic energy. The phenomenon of kinetic-energy driven superconductivity has also been observed experimentally for the cuprates~\cite{Deutscher,Gedik,Giannetti,Carbone}. However, the experimental trend is different in the sense that there is a transition close to optimal doping from BCS-type behaviour (for large doping values) to kinetic-energy driven superconductivity (for small doping values). In our previous calculations~\cite{buenemann2012b} we obtained similar behaviour for $U \gtrsim 13 |t|$, however the kinetic energy increase was always very small. The present results are more accurate and we observe that the kinetic energy change for overdoped systems is positive for $U \gtrsim 13 |t|$. Close to the critical doping the kinetic energy change is very small (of the order of $1$K). It might be below the accuracy of our method to determine whether it is positive or negative in this regime (by changing $l_c$ we could observe both for $l_c =11, 13, 15$). The lack of transition between the two regimes in Gutzwiller Wave Function has been remedied in the variational Monte Carlo method~\cite{Yokoyama1,Yokoyama2} by including in the projection also an additional Jastrow factor. 
This Jastrow factor is motivated by the form of a strong coupling expansion used to derive 
the $t$-$J$ model from the Hubbard model~\cite{Spalek1}. By including 
similar terms  in the present method it should be possible to observe a behaviour consistent with the 
experimental data. Work along this line is planned in the future.

\section{Summary and Outlook}\label{se4}
In summary, we have given a comprehensive derivation of a diagrammatic 
 variational method for the evaluation of superconducting states in two-dimensional 
 Hubbard models. Since most diagrams in our scheme are rather localised in real space
 we are able to evaluate them up to relatively large orders in the 
 expansion parameter $x$. For those  diagrams which are not localised, we 
 developed a resummation method that practically eliminates the numerical 
error in their real space evalution (e.g., we estimated this error to be of the order of $1$ K for the condensation energy). The remaining error of our method comes 
from the cutoff in the expansion in $x$. We have analyzed convergence of the 
condensation energy and correlated gap as a function of this cutoff. We have 
also studied the kinetic energy change upon condensation, a property that can
 be related to the experiment.

Our diagrammatic method is rather general and can be applied to various systems
 and situations. For the two-dimensional Hubbard model, it is still an open 
question 
   whether a Pomeranchuk and a superconducting state are competing or 
coexisting 
 near half filling. Also, the competition  or coexistence of these two phases
 with antiferromagnetic 
 order~\cite{kaczmarczyk2011} has not yet been studied.

It is further possible to apply the method to more complicated model 
systems, in particular to those, in which methods based on the Gutzwiller Approximation have provided valuable insights, e.g., periodic Anderson models~\cite{howczak2013,wysokinski2014}, 
multi-layer Hubbard models, or multi-band Hubbard 
models~\cite{buenemann1998,Zegrodnik3}.  Work in all these directions is 
in progress. Also the study of non-local interactions and/or
 correlations~\cite{abram2013} should be feasible in the future.  

\section*{Acknowledgements }
$ $ The work was supported by the Ministry of Science and Higher Education in
 Poland through the Iuventus Plus grant No. IP2012 017172 for the years 
2013-2015. JK also acknowledges support of the People Programme (Marie Curie Actions) of the European Union's Seventh Framework Programme (FP7/2007-2013) under REA grant agreement n$^{\rm o}$ [291734]
, as well as hospitality of the BTU Cottbus-Senftenberg where a large part of the work was performed. 
Access to the supercomputer located at ACMIN Centre of the AGH University of Science and Technology in Krak\'ow is also acknowledged.

\appendix 

\section{Treatment of local pairing}\label{app0}
In this Appendix, we explain how our diagrammatic formalism 
 can be applied if the local pairing condition~(\ref{456}) is not
 fulfilled (e.g. for superconducting phase with $s$-wave symmetry component). 

The basic idea remains the same as before, i.e, we aim to write 
 $\hat{P}_{\vecl}^{\dagger} \hat{P}_{\vecl}^{}$ in the form~(\ref{eq:1.690}) 
with an operator $\hat{d}^{\rm HF}_{\vecl}$ that ensures the vanishing
 of all Hartree bubbles at internal vertices. Now, however, 
 we need to cancel normal as well as anomalous Hartree bubbles.
 This requires the use of the more general local 
 correlation operator 
 \begin{equation}
\hat{P}_{\veci}=\sum_{\Gamma}\lambda_{\Gamma}
|\Gamma \rangle_{\veci\,\veci}\! \langle \Gamma |
+\lambda_B (|d \rangle_{\veci\,\veci}\! \langle \emptyset |+{\rm h.c.})\;.
\end{equation}   
Here we have already assumed that 
\begin{equation}\label{ure}
\Delta_0\equiv \langle \hat{\Delta}_{\vecl} \rangle_0\;\;\;\;
({\rm with}\;\; \hat{\Delta}_{\vecl} \equiv  \hat{c}_{\vecl,\downarrow}\hat{c}_{\vecl,\uparrow} )
\end{equation}   
is real which allows us to also work with a real parameter $\lambda_B$. 
 The following considerations can be readily generalised 
 in the case of a complex amplitude $\Delta_0$.

It will be useful to generalise 
 the  Hartree Fock operators~(\ref{4565}). Let 
\begin{equation}
\hat{O}_{\vecl}=\hat{\alpha}_{\vecl,1}\dots \hat{\alpha}_{\vecl,n}
\end{equation} 
 be an arbitrary operator on site $\vecl$ where 
$\hat{\alpha}_{\vecl,i}$ may be 
 creation or annihilation operators. Then, we want  
\begin{equation}
\hat{O}_{\vecl}^{\rm HF}=\hat{O}_{\vecl}-[\hat{O}_{\vecl}]^{\rm HF}
\end{equation}
 to create the same diagrams as 
 $\hat{O}_{\vecl}$
apart from those with Hartree bubbles at site $\vecl$.   
 This is achieved if we define $[\hat{O}_{\vecl}]^{\rm HF}$  recursively as
\begin{eqnarray} \label{hfop} 
&&\left[  
\hat{\alpha}_{\vecl,1}\ldots \hat{\alpha}_{\vecl,n}\right]^{\rm HF} 
\equiv\left \langle 
\hat{\alpha}_{\vecl,1}\ldots \hat{\alpha}_{\vecl,n} 
\right\rangle_{0}\\\nonumber 
&+&\mathop{{\sum}'}_{\{\gamma_1,...,\gamma_{n}\}=0}^{1} 
(-1)^{f_{\rm s}(\{\gamma_{i}\})} 
\left\{ 
\left( 
\prod_{\ell=1}^{n}\hat{\alpha}_{\vecl,\ell}^{\gamma_{\ell}} 
\right)- 
\left[ 
\prod_{\ell=1}^{n}\hat{\alpha}_{\vecl,\ell}^{\gamma_{\ell}} 
\right]^{\rm HF} 
\right\} \\\nonumber 
&&\times
\left\langle 
\prod_{\ell=1}^{n}\hat{\alpha}_{\vecl,\ell}^{1-\gamma_{\ell}} 
\right\rangle_{0}
\end{eqnarray}  
with  
\begin{equation} \label{8.70}
 f_{\rm s}(\{\gamma_{i}\})\equiv
\sum_{\ell=1}^{n}\left(\ell-\frac{1}{2}\right)\gamma_{\ell}  \; . 	 
\end{equation}
The prime in~(\ref{hfop}) indicates that
\begin{equation}
 2\leq \sum_{\ell=1}^{n}\gamma_{\ell}\leq n-2
\end{equation}
 has to be even (odd) if $n$ is even (odd). Due to this 
 summation restriction we find
\begin{eqnarray} \label{8.80}
\left[ \hat{\alpha}_{\vecl,i}  \hat{\alpha}_{\vecl,j} \right]^{\rm HF}&=&
\langle \hat{\alpha}_{\vecl,i}\hat{\alpha}_{\vecl,j}\rangle_{0} \;,\\
  \left[ \hat{\alpha}_{\vecl,i} \right]^{\rm HF}&=&0\;.
\end{eqnarray} 
 Note that this recursive definition is quite general and covers 
 systems with an arbitrary number of orbitals and local density matrices
 $\langle \hat{\alpha}_{\veci,i}\hat{\alpha}_{\veci,j}\rangle_0$. 
In the case of our single-band model with local pairing~(\ref{ure}), 
 we find, e.g.,
\begin{eqnarray}\label{axd}
 \hat{d}^{\rm HF}_{\vecl}&=&\hat{d}_{\vecl}
-[\hat{c}_{\vecl,\uparrow}^{\dagger}\hat{c}_{\vecl,\uparrow}
\hat{c}_{\vecl,\downarrow}^{\dagger}\hat{c}_{\vecl,\downarrow}]^{\rm HF}\\\nonumber
&=&\hat{d}_{\vecl}-n_0(\hat{n}_{\vecl,\uparrow}+\hat{n}_{\vecl,\downarrow})
-\Delta_0(\hat{\Delta}_{\vecl}+\hat{\Delta}^{\dagger}_{\vecl})+
n^2_0+\Delta^2_0
\end{eqnarray} 
 for the Hartree Fock operator in Eq.~(\ref{eq:1.6}). 
Another example which will be relevant for the evaluation 
 of hopping expectation values is
\begin{equation}
[\hat{c}_{\vecl,\uparrow}^{\dagger}\hat{c}_{\vecl,\uparrow}\hat{c}_{\vecl,\downarrow}]^{\rm HF}
=n_0\hat{c}_{\vecl,\downarrow}+\Delta_0\hat{c}_{\vecl,\uparrow}^{\dagger}\;.
\end{equation}

Together with~(\ref{axd}), the operator equation~(\ref{eq:1.6}). 
 leads to
\begin{eqnarray}\label{rqw}
 \lambda^2_d+\lambda^2_B&=&1+(1-n_0)^2x+\Delta^2_0x\;,\\
\lambda^2_{\sigma}&=&1-n_0(1-n_0)x+\Delta^2_0x\;, \\
 \lambda^2_{\emptyset}+\lambda^2_B&=&1+n_0^2x+\Delta^2_0 x\;,\\\label{rqw4}
\lambda_B(\lambda_d+\lambda_{\emptyset})&=&-x \Delta_0\;.
\end{eqnarray}
This set of equations determines, like in the case without local pairing, all 
parameters $\lambda_{\Gamma}$, $\lambda_{B}$ as a function 
of $x$. 

To calculate the expectation value of a local doble occupancy, 
we use the relation
\begin{eqnarray}
\hat{P}_{\veci}^{\dagger} \hat{d}_{\veci} \hat{P}_{\veci}^{}
&=&( \lambda^2_d+\lambda^2_B) \hat{d}^{\rm HF}_{\veci}\\\nonumber
&&+[(\lambda^2_d+\lambda^2_B)n_0-\lambda^2_B]
(\hat{n}^{\rm HF}_{\veci,\uparrow}+\hat{n}^{\rm HF}_{\veci,\downarrow})\\\nonumber
&&+[(\lambda^2_d+\lambda^2_B)\Delta_0-\lambda_B\lambda_d]
(\hat{\Delta}^{\rm HF}_{\veci}+{\rm h.c.})+\bar{d}_0
\end{eqnarray}
with 
\begin{equation}
\hat{\Delta}^{\rm HF}_{\veci}\equiv \hat{\Delta}_{\veci}-\Delta_0
\end{equation}
and
\begin{equation}
\bar{d}_0\equiv ( \lambda^2_d+\lambda^2_B)(n_0^2+\Delta_0^2)
+\lambda^2_B(1-2n_0)+2\lambda_B\lambda_d\Delta_0\;.
\end{equation}
This leads to 
\begin{eqnarray}
&&\langle \hat{d}_{\veci} \rangle_{\rm G}=\bar{d}_0
+2[(\lambda^2_d+\lambda^2_B)n_0-\lambda^2_B]I^{(2)}\\\nonumber
&&+( \lambda^2_d+\lambda^2_B-x\bar{d}_0 )I^{(4)}
+2[(\lambda^2_d+\lambda^2_B)\Delta_0-\lambda_B\lambda_d]I^{(2)}_{\rm s}
,
\end{eqnarray}
where we have introduced the (anomalous) diagrammatic sum
\begin{equation}
I^{(2)}_{\rm s}\equiv\sum_{k=0}^{\infty}\frac{x^k}{k!}
\sum_{\vecl_1,\ldots, \vecl_k}
\bigl\langle 
\hat{\Delta}^{\rm HF}_{\veci}
\hat{d}^{\rm HF}_{\vecl_1,\ldots,\vecl_k}
\bigr\rangle^{\rm c}_{0}\;.
\end{equation}

For the evaluation of a hopping expectation value we 
expand~(\ref{zxc}) , as in 
Eq.~(\ref{eq:1.10b}),
\begin{eqnarray}\nonumber
\widetilde{c}_{\veci,\sigma}^{\dagger}&=&
q \hat{c}^{\dagger}_{\veci,\sigma}+  \bar{q} \hat{c}^{}_{\veci,\bar{\sigma}}
+\alpha\Big ( \hat{c}^{\dagger}_{\veci,\sigma}  \hat{n}_{\veci,\bar{\sigma}}
-[ \hat{c}^{\dagger}_{\veci,\sigma}  \hat{n}_{\veci,\bar{\sigma}}]^{\rm HF}
\Big )\\
&&+\bar{\alpha} \Big ( \hat{c}^{}_{\veci,\bar{\sigma}}  \hat{n}_{\veci,\sigma}
-[ \hat{c}^{}_{\veci,\bar{\sigma}}  \hat{n}_{\veci,\sigma}]^{\rm HF}\Big )\\\nonumber
\widetilde{c}_{\veci,\sigma}^{}&=&
q \hat{c}^{}_{\veci,\sigma}+  \bar{q} \hat{c}^{\dagger}_{\veci,\bar{\sigma}}
+\alpha\Big ( \hat{c}^{}_{\veci,\sigma}  \hat{n}_{\veci,\bar{\sigma}}
-[ \hat{c}^{\dagger}_{\veci,\sigma}  \hat{n}_{\veci,\bar{\sigma}}]^{\rm HF}
\Big )\\
&&+\bar{\alpha} \Big ( \hat{c}^{\dagger}_{\veci,\bar{\sigma}}  \hat{n}_{\veci,\sigma}
-[ \hat{c}^{\dagger}_{\veci,\bar{\sigma}}  \hat{n}_{\veci,\sigma}]^{\rm HF}\Big )
\end{eqnarray}
with 
\begin{eqnarray}
q&\equiv&\lambda_1[\lambda_dn_0+\lambda_{\emptyset}(1-n_0)
+2\lambda_B \Delta_0]\;,\\
\bar{q}&\equiv&\lambda_1[(\lambda_d-\lambda_{\emptyset}) \Delta_0
+\lambda_{B}(1-2n_0)
+2\lambda_B \Delta_0]\;,\\
\alpha&\equiv&\lambda_1(\lambda_d-\lambda_\emptyset)\;,\\
\bar{\alpha}&\equiv&-2\lambda_1 \lambda_B\;.
\end{eqnarray}
Using the above equations, it is now a straightforward task 
 to write down the somewhat lengthy expressions for 
 $\langle \hat{c}^{\dagger}_{\veci,\sigma} \hat{c}_{\vecj,\sigma} \rangle_{\rm G}$
 and $\langle \hat{c}^{\dagger}_{\veci,\uparrow} \hat{c}^{\dagger}_{\vecj,\downarrow}\rangle_{\rm G}$.

\section{The linked-cluster theorem}\label{app1}
In order to apply the linked-cluster theorem in~(\ref{eq:1.690}) we need to 
lift the summation restrictions~(\ref{sre}) in 
Eqs.~(\ref{eq:1.9})-(\ref{eq:1.9c}). As we shall explain in this Appendix, the 
summation restriction can be lifted without generating additional terms. 

We consider a diagram $D$ in which the two operators 
$\hat{c}_{\vecl,\sigma}$ and  
$\hat{c}_{\vecl',\sigma}$ from the two internal vertices $\vecl$, $\vecl'$ 
 have a contraction with two other operators 
$\hat{\alpha}_1$, $\hat{\alpha}_2$, respectively.
 The operators  $\hat{\alpha_i}$ need not to be specified, i.e., they
 could be creation or annihilation operators and may belong to internal 
 as well as external vertices. The diagram $D$ results as one term 
in the evaluation of 
\begin{equation}\label{yds}
\langle \hat{c}_{\vecl',\sigma}  \hat{c}_{\vecl,\sigma} 
\hat{\alpha}_1, \hat{\alpha}_2 \hat{O}_{\rm rest}  \rangle_0
\end{equation}
by means of Wicks theorem where $\hat{O}_{\rm rest}$  contains 
all other operators which appear in $D$. Hence, we can 
 write $D$ as
  \begin{equation}\label{yds2}
D=\langle \hat{c}_{\vecl,\sigma} 
\hat{\alpha}_1 \rangle_0
\langle \hat{c}_{\vecl',\sigma}  \hat{\alpha}_2  \rangle_0
 D_{\rm rest}\;.
\end{equation} 
Another contribution from~(\ref{yds}), however, is 
 \begin{equation}\label{yds2b}
D'=-\langle \hat{c}_{\vecl,\sigma} 
\hat{\alpha}_2 \rangle_0
\langle \hat{c}_{\vecl',\sigma}  \hat{\alpha}_1  \rangle_0
 D_{\rm rest}\;.
\end{equation} 
If $\vecl=\vecl'$, we have $D+D'=0$, i.e., both diagrams cancel each other. 
 This explains why the summation restriction for the internal vertices can 
 be lifted. The same  arguments work if one of the
 two considered sites belongs to an external vertex, since 
  all four 
operators $\hat{c}^{(\dagger)}_{\vecl,\sigma}$ generate contractions 
 for each internal vertex. 

Note that after the application of the linked-cluster theorem, one must 
 {\sl not}  reintroduce the summation restrictions. For instance, the 
 diagram $D$ could be fully connected while in $D'\equiv D'_{\rm c} D'_{\rm dc}$ 
 the factor $ D'_{\rm dc}$ is cancelled by the norm. In such a case,
 the sum $D+D'_{\rm dc}$ which results after the application of the 
linked-cluster theorem will, in general, not be zero even for $\vecl=\vecl'$.

\section{Minimisation with respect to $|\Psi_0\rangle$}\label{app5}
\subsection{Energy functional without long-range diagrams}
If we ignore the non-locality of long-range diagrams, our Lagrange functional
 depends on $x$ and $|\Psi_0\rangle$ where $|\Psi_0\rangle$ enters the functional  
only via the elements of the single-particle density matrix   
 \begin{equation}
\tilde{\rho}^{\rm s} \equiv \left( 
\begin{array}{cc}
\tilde{\rho}& \tilde{\Delta}\\
\tilde{\Delta}^{\dagger}& 1-\tilde{\rho}
\end{array}
 \right)  \;.
\end{equation} 
Here we introduced
 \begin{eqnarray}
\rho_{(\veci \sigma),(\vecj \sigma)}&\equiv& \langle \hat{c}_{\vecj \sigma}^{\dagger} \hat{c}_{\vecj \sigma}  \rangle_{0}\\
\Delta_{(\veci \sigma),(\vecj \bar{\sigma})}&\equiv&
 \langle \hat{c}_{\vecj \bar{\sigma}} \hat{c}_{\vecj \sigma}  \rangle_{0}\;.
\end{eqnarray} 
Since $\tilde{\rho}^{\rm s}$ is derived from a single-particle product state
 it has to obey the constraint 
$\tilde{\rho}^{\rm s}\cdot \tilde{\rho}^{\rm s}=\tilde{\rho}^{\rm s}$\;.
Like in the paramagnetic case we implement this constraint in the minimisation with 
 respect to $ \tilde{\rho}^{\rm s}$ by means of Lagrange parameters, 
see Refs.~\cite{seibold2008b,buenemann2012c}. 
This leads to the 
 self-consistent single-particle problem 
 introduced in Eqs.~(\ref{eq:iou1}), (\ref{rde})-(\ref{eq:iouF})

\subsection{Energy functional with long-range diagrams}
When we evaluate the long-range diagrams as described in Section~\ref{app4}, we obtain 
 an energy functional that does not only depend on the elements 
of $\tilde{\rho}^{\rm s}$  in real space, i.e., the 
lines $X^{\alpha}_{\veci,\vecj}$. It also depends on the `higher-order' lines 
   $Y^{\alpha,\alpha'}_{\veci,\vecj}$ and  $Z^{\alpha_1,\alpha_2,\alpha_3}_{\veci,\vecj}$
 which, on the other hand, 
are determined by the elements of  $\tilde{\rho}^{\rm s}$ in 
 momentum space, i.e., the distributions $X_{\veck}^{\alpha}$. Therefore, 
 the great canonical potential has the form
\begin{eqnarray}\label{tfr}
&&\mathcal{F}(\{ P_{\veci,\vecj}\},\{ S_{\veci,\vecj}\};
\{X_{\veck}^{\alpha} \})\\\nonumber
&&\equiv\mathcal{\bar{F}}\big(\{ P_{\veci,\vecj}\},\{ S_{\veci,\vecj}\};\\\nonumber
&&\;\;\;\;\;\;\;\;\;\{Y^{\alpha,\alpha'}_{\veci,\vecj}[\{  X_{\veck}^{\alpha} \}]\},
\{Z^{\alpha_1,\alpha_2,\alpha_3}_{\veci,\vecj}[\{  X_{\veck}^{\alpha} \}]\}
\big )\;.
\end{eqnarray} 
Note that the lines $P_{\veci,\vecj}$, $S_{\veci,\vecj}$ enter the functional also 
 via the evaluation of long-range diagrams, see Eqs.~(\ref{34r1}), (\ref{34r2}),
 (\ref{38r2}), (\ref{3458}). 
 The minimisation of $\mathcal{F}$ with respect to  $P_{\veci,\vecj}$, $S_{\veci,\vecj}$, and 
 $X_{\veck}^{\alpha}$ leads to 
 \begin{equation}\label{6re}
\hat{H}_0^{\rm eff}|\Psi_0\rangle=E_0  |\Psi_0\rangle
\end{equation}
where 
\begin{eqnarray}
\hat{H}_0^{\rm eff} &=&
\sum_{\veci,\vecj,\sigma}t^{\rm eff}_{\veci,\vecj}
\hat{c}_{\veci,\sigma}^{\dagger}\hat{c}_{\vecj,\sigma}^{\phantom{\dagger}}
+ \sum_{\veci \neq \vecj} \bigl( \Delta^{\rm eff}_{\veci,\vecj}
\hat{c}_{\veci,\ua}^{\dagger}\hat{c}_{\vecj,\da}^{\dagger} +
{\rm h.c.} \bigr) \\
&&+\sum_{\veck,\sigma} \epsilon_{\veck} \hat{n}_{\veck,\sigma} 
+\left(\sum_{\veck} \Delta_{\veck} \hat{c}^{\dagger}_{\veck,\uparrow} 
\hat{c}^{\dagger}_{-\veck,\downarrow}  +
{\rm h.c.}\right)
\end{eqnarray}
and 
\begin{eqnarray}
t^{\rm eff}_{\veci,\vecj} &=&
\frac{\partial \mathcal{F}}{\partial P_{\veci,\vecj}}\;,\quad \Delta^{\rm eff}_{\veci,\vecj} =
\frac{\partial \mathcal{F}}{\partial S_{\veci,\vecj}} \;,
\\
\epsilon_{\veck} &=&\frac{\partial \mathcal{F}}{\partial X^1_{\veck}}\;,\quad
\Delta_{\veck} =\frac{\partial \mathcal{F}}{\partial X^2_{\veck}}\;.
\end{eqnarray}
Equation~(\ref{tfr}) then yields
\begin{eqnarray}
 \epsilon_{\veck}&=&\sum_{\alpha}(r^{1 ,\alpha}_{\veck}+r^{\alpha,1}_{\veck})X^{\alpha}_{\veck}\\\nonumber
&&+\sum_{\alpha,\alpha'}(v^{1 ,\alpha ,\alpha'}_{\veck}+v^{\alpha, 1, \alpha'}_{\veck}+v^{\alpha, \alpha', 1}_{\veck})X^{\alpha}_{\veck}
 X^{\alpha'}_{\veck}\;,\\ 
\Delta_{\veck}&=&\sum_{\alpha}(r^{2 ,\alpha}_{\veck}+r^{\alpha, 2}_{\veck})X^{\alpha}_{\veck}\\\nonumber
&&+\sum_{\alpha,\alpha'}(v^{2, \alpha ,\alpha'}_{\veck}+v^{\alpha, 2, \alpha'}_{\veck}+v^{\alpha, \alpha', 2}_{\veck})X^{\alpha}_{\veck}
 X^{\alpha'}_{\veck}\;.
\end{eqnarray}
where 
\begin{eqnarray}
r^{\alpha, \alpha'}_{\veck}&\equiv& \frac{1}{L}\sum_{\veci,\vecj}e^{{\rm i}\veck(\vecj-\veci)}
\frac{\partial }
{\partial Y_{{\veci,\vecj}}^{\alpha,\alpha'}}\bar{\mathcal{F}}\;,\\
v^{\alpha_1, \alpha_2, \alpha_3}_{\veck}&\equiv& \frac{1}{L}\sum_{\veci,\vecj}e^{{\rm i}\veck(\vecj-\veci)}
\frac{\partial }
{\partial Z_{{\veci,\vecj}}^{\alpha_1, \alpha_2, \alpha_3}}\bar{\mathcal{F}}\;.
\end{eqnarray} 
\bibliographystyle{pss}
\bibliography{bib5}

\end{document}